\documentclass[%
 aip,
 amsmath,amssymb,
preprint,
]{revtex4-1}

\usepackage{graphicx}
\usepackage{dcolumn}
\usepackage{bm}
\usepackage[utf8]{inputenc}
\usepackage[T1]{fontenc}
\usepackage{mathptmx}

\begin{document}


\title[Temperature dependence of the Seebeck coefficient of epitaxial \hbox{$\beta$-Ga$_2$O$_3$} thin films]{Temperature dependence of the Seebeck coefficient of epitaxial \hbox{$\beta$-Ga$_2$O$_3$} thin films}

\author{Johannes Boy}
\email[]{boy@physik.hu-berlin.de}
\affiliation{Novel Materials Group, Humboldt-Universität zu Berlin, Newtonstraße 15, 12489 Berlin, Germany}
\author{Martin Handwerg}
\affiliation{Novel Materials Group, Humboldt-Universität zu Berlin, Newtonstraße 15, 12489 Berlin, Germany}
\author{Robin Ahrling}
\affiliation{Novel Materials Group, Humboldt-Universität zu Berlin, Newtonstraße 15, 12489 Berlin, Germany}
\author{R\"udiger Mitdank}
\affiliation{Novel Materials Group, Humboldt-Universität zu Berlin, Newtonstraße 15, 12489 Berlin, Germany}
\author{G\"unter Wagner}
\affiliation{Leibniz-Institut f\"ur Kristallz\"uchtung, Max-Born-Strasse 2, 12489 Berlin, Germany}
\author{Zbigniew Galazka}
\affiliation{Leibniz-Institut f\"ur Kristallz\"uchtung, Max-Born-Strasse 2, 12489 Berlin, Germany}
\author{Saskia F. Fischer}
\affiliation{Novel Materials Group, Humboldt-Universität zu Berlin, Newtonstraße 15, 12489 Berlin, Germany}

\date{\today}

\begin{abstract}
The temperature dependence of the Seebeck coefficient of homoepitaxial metal organic vapor phase (MOVPE) grown, silicon doped \hbox{$\beta$-Ga$_2$O$_3$} thin films was measured relative to aluminum. For room temperature we found the relative Seebeck coefficient of $S_{\beta\text{-Ga}_2\text{O}_3\text{-Al}}=(-300\pm20)\;\mu$V/K. At high bath temperatures $T>240$ K, the scattering is determined by electron-phonon-interaction. At lower bath temperatures between $T=100$ K and $T=300$ K, an increase in the magnitude of the Seebeck coefficient is explained in the frame of Strattons formula. The influence of the different scattering mechanisms on the magnitude of the Seebeck coefficient is discussed and compared with Hall measurement results.
\end{abstract}

\maketitle

\section{Introduction}
In the past years, \hbox{$\beta$-Ga$_2$O$_3$} crystals and thin films have proved to be promising materials for high power devices \cite{Step_Rev,Suzuki,Green2016,Chabak2016,Galazka2018}. However, one drawback is the energy-dissipation, which enhances Joule heating, due to the low thermal conductivity\cite{HandwergLambda1,HandwergLambda2}. One approach could be a direct cooling using the Peltier effect.
For this purpose values of the Seebeck coefficient of \hbox{$\beta$-Ga$_2$O$_3$} must be known.\\
\hbox{$\beta$-Ga$_2$O$_3$} is a transparent material, with a high band gap $E_\text{g}=4.7 - 4.9$ eV at room temperature\cite{Tippins,Lorenz,Orita,Peelaers,Janowitz}.
The majority charge carrier type is n-type and the effective mass has been experimentally determined to be in the order of $m^*=0.25 - 0.28$ free electron masses \cite{Kang,Mohamed,Janowitz}. \hbox{$\beta$-Ga$_2$O$_3$} has been intensively studied in terms of charge carrier transport with a maximum mobility so far being $\mu=153$ cm$^2$/Vs\cite{Galazka2014} at room temperature. However, electrical\cite{Ahrling2018,Mitdank2014} and thermal\cite{HandwergLambda1,HandwergLambda2} conductivity studies remain to be completed by thermoelectric measurements.\\
In this work we implement a micro lab, based on metallic lines on the homoepitaxially MOVPE grown (100) silicon doped $\beta$-Ga$_2$O$_3$ thin films and perform temperature-dependent Seebeck measurements between $T=100$ K and $T=300$ K. We compare the results with calculated room temperature Seebeck coefficients based on Hall charge carrier density by using Stratton's formula\cite{Stratton}.\\

\section{Experimental Details}
The thin films have been grown on substrates prepared from Mg-doped, electrically insulating bulk \hbox{$\beta$-Ga$_2$O$_3$} single crystals, that were grown along the [010]-direction by the Czochralski method\cite{Galazka2014,Galazka2016}. The substrates with (100)-orientation have been cut with a 6$^\circ$ off-orientation to reduce island growth\cite{Galazka2010,Galazka2016} and increase the structural quality of the thin films. The MOVPE process used trimethylgallium and pure oxygen as precursors. Silicon doping has been realized by tetraethylorthosilicate. The substrate temperature was between 750$^\circ$C and 850$^\circ$C and the chamber pressure between 5 and 100 mbar during growth\cite{Wagner2014}.\\
\begin{table}
\caption{Thin film thickness $d$, charge carrier density $n$ and mobility $\mu$ for the investigated \hbox{$\beta$-Ga$_2$O$_3$} thin film samples at room temperature and 100 K. We have obtained doping densities in the range of $N_\text{D}=2.1\cdot10^{18}$ cm$^{-3}$ and $N_\text{D}=2.2\cdot10^{18}$ cm$^{-3}$ for the high and low mobility sample, respectively, by fits of the temperature dependent charge carrier densities.}\label{MatPar}
\begin{ruledtabular}
\begin{tabular}{ccc} 
\hbox{$\beta$-Ga$_2$O$_3$} thin film & sample 1 & sample 2\\
 & high mobility & low mobility\\ 
\hline 
$d$ [nm] & 185 & 212 \\ 
$n(T=300\;\text{K})$ [cm$^{-3}$]& $(5.5\pm0.1)\cdot10^{17}$ & $(6.2\pm0.1)\cdot10^{17}$ \\ 
$\mu(T=300\;\text{K})$ [cm$^2$/Vs] & $103\pm1$ & $29\pm1$ \\ 
$n(T=100\;\text{K})$ [cm$^{-3}$]& $(1.6\pm0.1)\cdot10^{17}$ & $(2.1\pm0.1)\cdot10^{17}$ \\  
$\mu(T=100\;\text{K})$ [cm$^2$/Vs] & $233\pm1$ & $48\pm1$ \\ 
\end{tabular} 
\end{ruledtabular}
\end{table}
The material parameters of the samples are listed in table \ref{MatPar}. Both \hbox{$\beta$-Ga$_2$O$_3$} thin films have the same magnitude of charge carrier densities but a rather big difference in mobility.\\
The micro labs have been manufactured by standard photolithography and magnetron sputtering of titanium (7 nm) and gold (35 nm) after cleaning with acetone, isopropanol and subsequent drying. The metal lines of the micro lab are isolated due to a Schottky contact relative to the \hbox{$\beta$-Ga$_2$O$_3$} thin film. Ohmic contacts with the \hbox{$\beta$-Ga$_2$O$_3$} thin film were achieved by direct wedge bonding with an Al/Si-wire (99/1 \%) on the deposited metal structure\cite{Ahrling2018}, creating point contacts. To keep some parts of the micro lab isolated relative to the thin film, the electrical contacts were prepared by attaching gold wire with indium to the Ti/Au metal lines.\\
\begin{figure}
\includegraphics[scale=1]{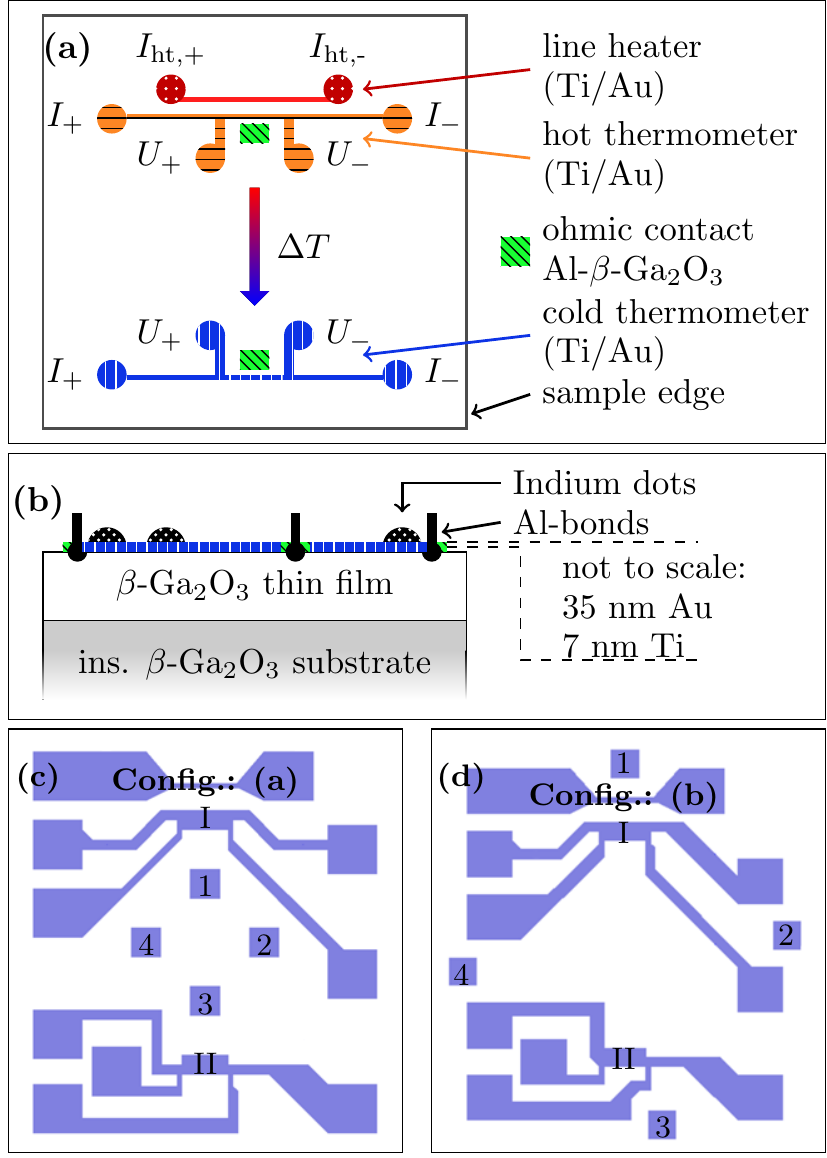} 
\caption{\label{Scheme} Figure (a) gives a schematic overview of the function of the Seebeck micro lab. A two point conductor at the top of the scheme (red with white dots) serves as a line heater to create a temperature gradient $\Delta T$ across the sample. The temperature difference is being measured by the change of the four point resistance in the thermometer lines at the bottom of the scheme (cold thermometer, blue with white vertical lines) and below the line heater (hot thermometer, orange with horizontal black lines). The little squares (green with diagonal black lines) mark the ohmic contacts, which are used to measure the thermo voltage ($U_\text{th}$).  Figure (b) illustrates the cross-section of the samples, see text for details. Figure (c) and (d) show a scheme of the photolithographic created measurement configurations (a) and (b), respectively. I and II mark the ohmic thermo voltage contacts, 1-4 mark the ohmic four point measurement contacts (not shown in (a)).}
\end{figure}
Figure \ref{Scheme} shows the micro lab, that allows the measurement of the Seebeck coefficient, charge carrier density and conductivity.\\
Figure \ref{Scheme} (a) displays a scheme of the micro lab. It consists of a two-point line heater at the top, were an electrical current is imprinted to create Joule heating, resulting in a temperature gradient $\Delta T$ across the sample.
Two four-point metal lines close to (hot) and far from (cold) the heater line serve as thermometers. The temperature dependent measurement of their resistances follow the Bloch-Gr\"uneisen-law\cite{Tritt} and can be used to calculate the temperature gradient in the area of the four-point resistances.\\
In this area, Al-\hbox{$\beta$-Ga$_2$O$_3$} ohmic contacts were placed to measure the thermo voltage.\\
The experimental procedure involves a measurement of the thermo voltage for as long as it takes to stabilize the temperature gradient across the sample. Afterwards, the four-point resistances of the thermometers are measured. This procedure is repeated for several heating currents before the bath temperature is changed in intervals of 10 K.\\
Figure \ref{Scheme} (b) illustrates a cross-sectional view of the samples. The Mg-doped electrically insulating \hbox{$\beta$-Ga$_2$O$_3$} substrate has been used to grow the Si-doped \hbox{$\beta$-Ga$_2$O$_3$} thin film on top thereof. The Ti/Au metal lines of the micro lab have then been deposited on the surface of the thin film and electrical contacts have been prepared either by wedge-bonding with Al-wire (ohmic contact) or attachement of Au-wire with indium (Schottky contact).\\
Figure \ref{Scheme} (c) and (d) show two different designs (a) and (b) for electrical measurements in the 4-point van-der-Pauw configuration and differences in the 4-point thermometers. In figure \ref{Scheme} (c) the van-der-Pauw contacts (1-4) are located in the centre of the sample.
To check the validity of the measurement setup, an alternative measurement configuration has been developed (measurement configuration (b)).\\
The measurement configuration (b) is shown in figure \ref{Scheme} (d). Here, the van-der-Pauw contacts are close to the edge of the sample. 
The measurement configuration a has been used to investigate the \hbox{185 nm} thick high mobility sample and one \hbox{212 nm} thick low mobility sample. To verify that there is no influence because of the van-der-Pauw contacts, a second \hbox{212 nm} thick low mobility sample has been investigated with the measurement configuration (b).\\ 

\section{Measurement Results}
The thermo voltage $U_\text{th}$ was measured as a function of temperature gradient $\Delta T$ for several bath temperatures and \hbox{$\beta$-Ga$_2$O$_3$} thin films.
\begin{figure}
\includegraphics[scale=1]{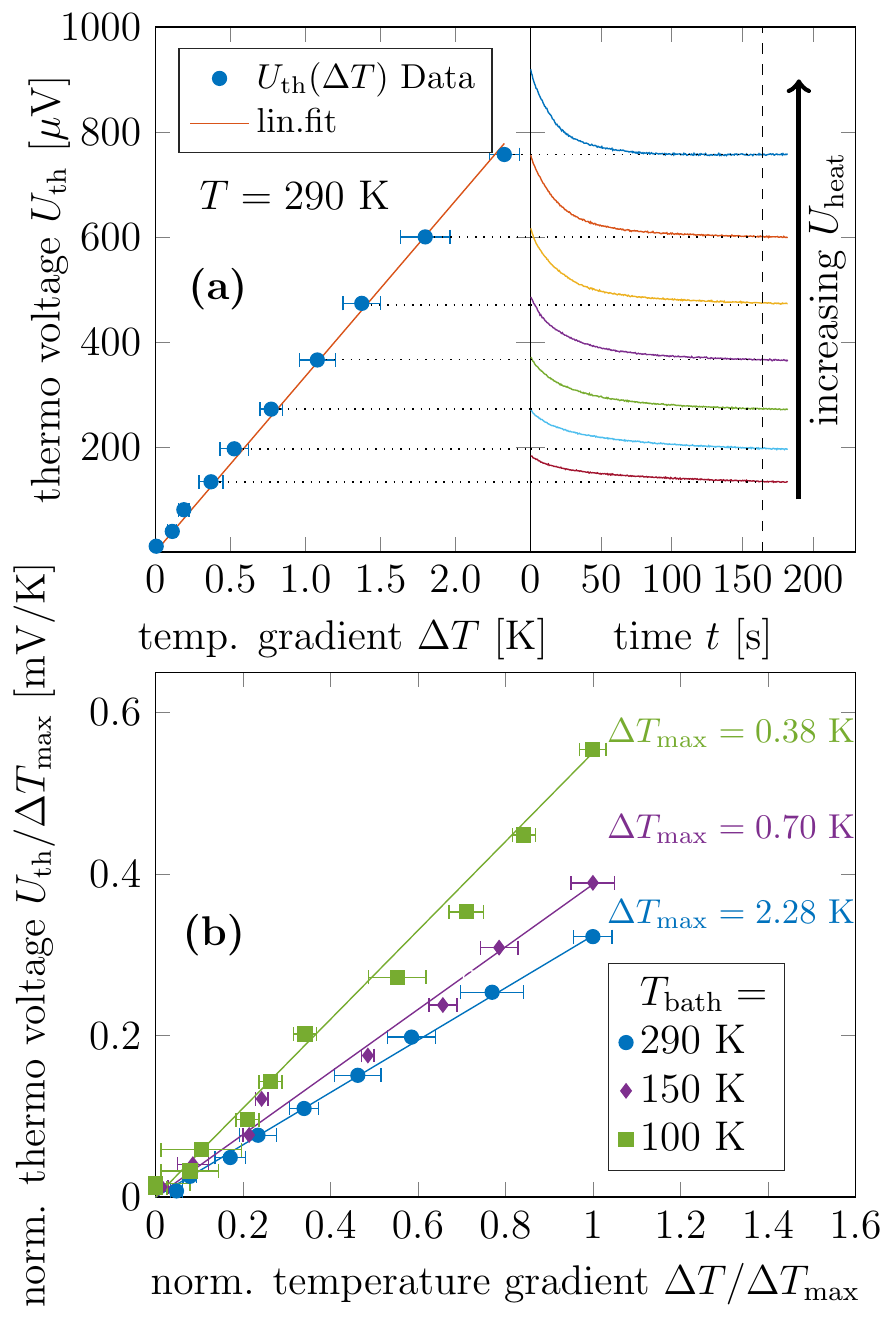} 
\caption{\label{dUth_dT} Figure (a): The thermo voltage as a function of time (right hand side) and temperature gradient (left hand side) for the 185 nm thick MOVPE \hbox{$\beta$-Ga$_2$O$_3$} thin film with high mobility for a bath temperature of  290 K is shown. The measurement of the thermo voltage has been performed for 180 seconds after applying a heating voltage to the line heater. Only the last 10 \% of the measured data (marked by the vertical dashed line) have been used for the evaluation to make sure a stable temperature gradient has evolved. The left graph shows the average of the thermo voltage $U_\text{th}$ as a function of the temperature difference $\Delta T$ between the hot and the cold thermometer. The dotted lines are a guidance for the eye. A linear fit $U_\text{th}=-S\cdot\Delta T+U_\text{off}$ has been applied to determine the Seebeck coefficient $S$. (b) The measured thermo voltage $U_\text{th}$ normalized by the maximum of the temperature gradient for the same sample has been plotted as a function of the normalized temperature gradient. An offset $U_\text{os}<50\;\mu$V has been substracted from the plotted data. The corresponding bath temperatures are from bottom to top: \hbox{290 K}, \hbox{150 K} and \hbox{100 K}.}
\end{figure}
In the right graph of figure \ref{dUth_dT} (a) the measurements of the thermo voltage as a function of time for different temperature gradients can be seen. The measurements are performed for 180 seconds and the last 10 \% of the measured data (marked by the vertical dashed line) are used to evaluate the thermo voltage as a function of temperature gradient (left graph). Due to the weaker thermal conductivity and diffusivity of \hbox{$\beta$-Ga$_2$O$_3$} at higher temperatures\cite{HandwergLambda1,HandwergLambda2}, it takes longer to evolve a stable temperature gradient and thermo voltage as compared to lower temperatures.\\
The left graph of figure \ref{dUth_dT} (a) shows the thermo voltage as a function of the temperature gradient. These data can be well fitted with a linear equation, where the slope equals the Seebeck coefficient\\
\begin{equation}
S=-\frac{U_\text{th}}{\Delta T}.
\end{equation}
Figure \ref{dUth_dT} (b) shows the normalized thermo voltage as a function of normalized temperature gradient for bath temperatures of \hbox{290 K}, \hbox{150 K}, and \hbox{100 K} with substracted offsets for the same sample as figure \ref{dUth_dT} (a). The change of the Seebeck coefficient as a function of bath temperature can be observed by the change of the slope of the linear fits.\\ 
\begin{figure}
\includegraphics[scale=1]{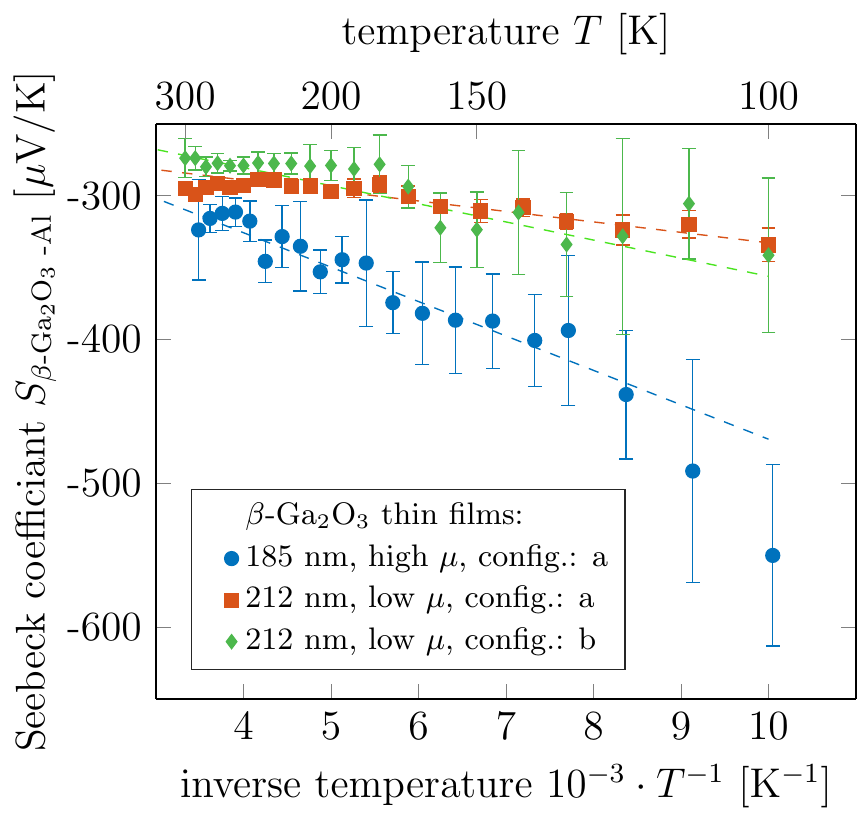} 
\caption{\label{S_1T} Seebeck coefficient $S$ as a function of the inverse Temperature $T^{-1}$ for the investigated \hbox{185 nm} thick high mobility sample 1a (blue circles) and 212 nm low mobility samples 2a (red squares) and 2b (green diamonds) relative to aluminum. A linear fit has been performed for temperatures above 130 K to investigate the Stratton formula \hbox{$S=-\frac{k_\text{B}}{e}\left(r+\frac{5}{2}-\frac{E_\text{F}-E_\text{C}}{k_\text{B}T}\right)$} for constant $E_\text{F}$ and $r$.}
\end{figure}
Figure \ref{S_1T} displays the measured Seebeck coefficients of different \hbox{$\beta$-Ga$_2$O$_3$} thin films as a function of inverse bath temperature between 100 K and 300 K. The Seebeck coefficients are in the range of $S=-280\;\mu$V/K to $S=-500\;\mu$V/K above 100 K. At room temperature, the Seebeck coefficients are $S=-(300\pm20)\;\mu$V/K for all investigated \hbox{$\beta$-Ga$_2$O$_3$} samples. The correction of the Seebeck coefficient due to the aluminum wire is less than $1\;\%$. The high mobility \hbox{$\beta$-Ga$_2$O$_3$} thin film shows a stronger increase of the Seebeck coefficient for lower temperatures than the low mobility \hbox{$\beta$-Ga$_2$O$_3$} thin films. We find, that the investigated measurement configuration has no detectable influence on the Seebeck coefficient. The lower magnitude of the Seebeck coefficient for the \hbox{212 nm} thick low mobility sample at low temperatures is expected to be due to different charge carrier scattering mechanisms.\\
These results are similar to those of other transparent conducting oxides like ZnO or In$_2$O$_3$, where Seebeck coefficients $S$ in the range from $S=-20\;\mu$V/K to $S=-500\;\mu$V/K have been measured\cite{Barasheed2013, Bierwagen2015}.

\section{Discussion}
The study of the temperature dependent Seebeck coefficient for \hbox{$\beta$-Ga$_2$O$_3$} thin films gives an insight into the scattering mechanisms within the material\cite{Young_Ss,Herring_STh,Ginley}. Here we discuss the obtained Seebeck coefficients and their description by the Stratton formula \cite{Stratton}:\\
\begin{equation}\label{StrattonForm}
S_\text{th}=-\frac{k_\text{B}}{e}\left(r+\frac{5}{2}+\eta\right),
\end{equation}
with the elemental charge e, the scattering parameter $r$, the reduced electron chemical potential \hbox{$\eta=(E_\text{C}-E_\text{F})/(k_\text{B}T)$} and the conduction band energy $E_\text{C}$, Fermi energy $E_\text{F}$ and Boltzmann constant $k_\text{B}$. The scattering parameter is related to the scattering mechanisms within the sample. A first approximation of formula (2) has been plotted in figure (3). A linear fit with constant $r=0.3\pm0.2$ and $E_\text{C}-E_\text{F}$ has been performed for $T>130$ K and plotted in figure \ref{S_1T} (dashed lines). For the \hbox{185 nm} thick high mobility sample $E_\text{C}-E_\text{F}\approx 24$ meV and for the \hbox{212 nm} thick low mobility samples $E_\text{C}-E_\text{F}\approx 13$ meV has been obtained. These different electron chemical potentials $E_\text{C}-E_\text{F}$ explain the different temperature dependencies but only give a rough estimation since $r=r(T)$ and $E_\text{F}=E_\text{F}(T)$.\\
To calculate theoretical values for the Seebeck coefficient at room temperature, we use the analytical expression after Nilsson \cite{Nilsson1973}, which interpolates the range between non degenerated and degenerated semiconductors. The reduced electron chemical potential $\eta$ is then given by\\
\begin{equation}
\eta=\frac{\ln\frac{n}{N_\text{C}}}{1-\left(\frac{n}{N_\text{C}}\right)^2}+\nu\left(1-\frac{1}{1+(0.24+1.08\nu)^2}\right),
\end{equation}
\begin{equation}
\nu=\left(\frac{3\sqrt{\pi}\frac{n}{N_\text{C}}}{4}\right)^{2/3},
\end{equation}
with $N_\text{C}$ beeing the effective density in the conduction band:\\
\begin{equation}
N_\text{C}=2\left(\frac{2\pi m^*k_\text{B}T}{h^2}\right)^{3/2}.
\end{equation}
With an effective mass of \hbox{$m^*=(0.23\pm0.02)\cdot$m$_\text{e}$}, which we obtained by fitting the measured temperature dependent charge carrier density with the neutrality equation, and a Seebeck scattering parameter of $r=-1/2$\cite{Ginley,Preissler2013,Sze}, which applies for electron-phonon scattering we calculated theoretical values using the Stratton formula \eqref{StrattonForm}. That calculation yields a value of \hbox{$S_\text{th}(T>$ 250 K$)=(-300\pm20)\;\mu$V/K},
\begin{figure}
\includegraphics[scale=1]{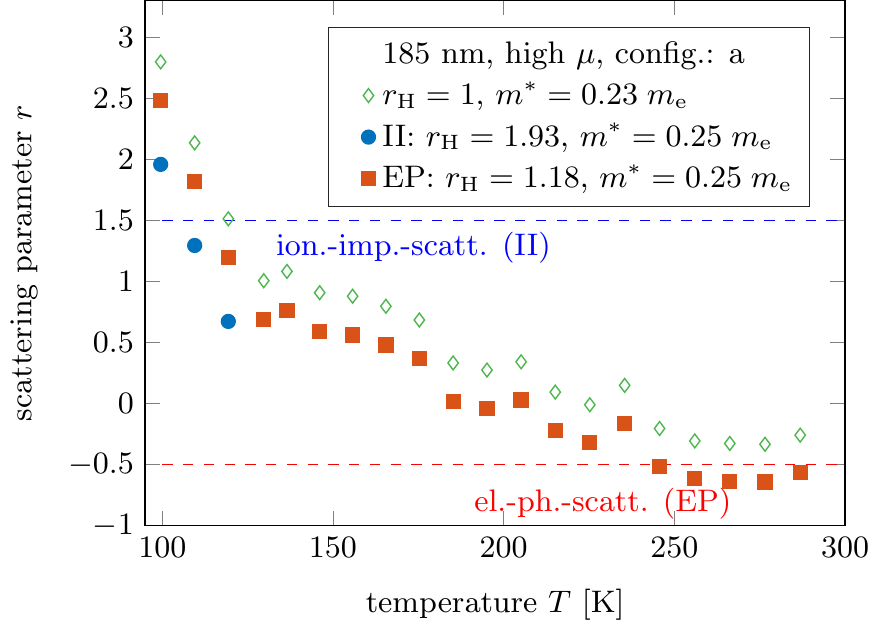} 
\caption{\label{r_T} Temperature dependent scattering parameters considering the measured Seebeck coefficients and measured charge carrier density for the investigated 185 nm thick high mobility sample. Various effective masses $m^*$ and Hall scattering parameters $r_\text{H}$ have been considered, using equation \eqref{StrattonForm}. For the open diamonds no correction of the calculated $n_\text{H}$ has been applied and \hbox{$m^*=0.23\;m_\text{e}$}. The blue dots and red triangles consider $m^*=0.25\;m_\text{e}$ as reported in the literature\cite{Janowitz} as well as Hall scattering factors for ionized impurity (II) scattering and electron-phonon (EP) interaction, respectively. Scattering parameters of $r=-0.5$ and $r=1.5$ are expected for EP and II interaction, respectively.}
\end{figure}
The measurement of the Seebeck coefficient and charge carrier density allows a calculation of the scattering parameter $r$ if the effective mass $m^*$ is known. One has to consider, that the measurement of the charge carrier density using Hall measurements is influenced by the scattering of the free electrical charges as well. This is usually described by the Hall scattering factor $r_\text{H}$\cite{Sze}, if the relaxation time $\tau$ can be expressed by $\tau\propto E^r$:
\begin{equation}
r_\text{H}=\Gamma(5/2-2r)\cdot \Gamma(5/2)/(\Gamma(5/2-r))^2=\mu_\text{H}/\mu,
\end{equation}
with $\Gamma(x)$ being the gamma function, $\mu_\text{H}$ the Hall mobility and $\mu$ the drift mobility.
In order to correctly calculate the general scattering parameter $r$ one needs to know the Hall scattering parameter $r_\text{H}$ or vice versa. Commonly for electron phonon (EP) scattering $r_\text{H}=1.18$ and for ionized impurity (II) scattering $r_\text{H}=1.93$ are assumed\cite{Sze}. This allows a calculation of the scattering parameter $r$, if we assume that only II or EP scattering dominate the charge transport. Figure \ref{r_T} shows the calculated $r$ as a function of bath temperature. The calculations were performed with equation \eqref{StrattonForm} considering the measured Seebeck coefficients and measured charge carrier densities, as well as various Hall scattering factors and effective masses for the 185 nm thick high mobility \hbox{$\beta$-Ga$_2$O$_3$} thin film.\\
For $r_\text{H}=1$ and $m^*=0.23\;m_\text{e}$ (open diamonds) no correction of the charge carrier density has been done and the effective mass has been used as obtained from temperature dependent charge carrier densities fits. If we consider Hall scattering factors\cite{Sze} of $r_\text{H}=1.93$ and $r_\text{H}=1.18$ for ionized impurity scattering and electron-phonon interaction, respectively, and assume an effective mass of $m^*=0.25\;m_\text{e}$ as reported in the literature\cite{Janowitz}, we obtain the blue dots and red squares, for II and EP scattering respectively. The obtained results near room temperatures are in agreement with the expected value of $r=-0.5$ for electron-phonon-scattering (figure \ref{r_T} lower dashed line) \cite{Ginley,Preissler2013,Sze}. For \hbox{$\beta$-Ga$_2$O$_3$} thin films electron-phonon-scattering is expected to be the dominant scattering mechanism at these temperatures\cite{Ahrling2018}. Here, the scattering mechanism dominating the charge transfer caused by electrical potential gradients is the same as caused by temperature gradients.\\
For low temperatures we obtain values close to \hbox{$r=3/2$} as expected for ionized-impurity scattering (upper dashed line). This is also in agreement with values reported in literature\cite{Ginley,Preissler2013,Sze}. Ionized impurity scattering has been reported\cite{Ahrling2018,Oishi2015} to become the dominant scattering mechanism in \hbox{$\beta$-Ga$_2$O$_3$} below 100 K, depending on layer thickness and doping. This explains why we obtain values close to 1.5 if we assume that the charge carrier scattering mechanisms influencing the mobility and the Seebeck effect are the same.\\
The Seebeck coefficients in \hbox{$\beta$-Ga$_2$O$_3$} thin films at room temperature are mainly dependent on the doping level, since electron phonon scattering is the dominant scattering mechanism, whereas at low temperatures the different dominant scattering mechanisms have a major impact on the magnitude of the Seebeck coefficient. These observations are consistent with previous studies on thermal\cite{HandwergLambda1,HandwergLambda2} and electrical\citep{Ahrling2018,Mitdank2014} conductivity where the dominant scattering mechanisms at room temperature are phonon-based. For applications at higher temperatures, the room temperature magnitude of the Seebeck coefficient gives an upper limit. Furthermore we can estimate that $S$ approaches a value of $2k_\text{B}/e$ in the intrinsic regime.\\

\section{Conclusion}
In conclusion, the temperature dependent Seebeck coefficient of homoepitaxial \hbox{$\beta$-Ga$_2$O$_3$} thin films can be explained by the Stratton formula. \hbox{$\beta$-Ga$_2$O$_3$} thin films have Seebeck coefficients relative to aluminum of \hbox{$S_{\beta\text{-Ga}_2\text{O}_3\text{-Al}}=(-300\pm20)\;\mu$V/K} at room temperature with an increase in magnitude at lower temperatures. This leads to a room temperature Peltier coefficient of \hbox{$\Pi\approx0.1$ V}.
The dependency of the Seebeck coefficent on the dominant scattering mechanism gives the possibility of Seebeck coefficient engineering by growing \hbox{$\beta$-Ga$_2$O$_3$} thin films with, for example, an increased concentration of neutral impurities or ionized impurities.

\section*{Acknowledgement}
This work was performed in the framework of GraFOx, a Leibniz-ScienceCampus partially funded by the Leibniz association and by the German Science Foundation (DFG-FI932/10-1 and DFG-FI932/11-1). The authors thank M. Kockert for fruitful scientific discussions.

\nocite{*}
\bibliography{mybibfile}

\end{document}